\documentclass[12pt]{article}

\usepackage[margin=1in]{geometry}
\usepackage{setspace}
\usepackage{graphicx}
\usepackage{refstyle}
\usepackage{caption}
\usepackage{subcaption}
\usepackage{multirow}
\usepackage{authblk}
\usepackage{color}

\begin{document}

\title{Graphene Nanoribbons Under Axial Compressive and Point Tensile Stresses}
\author[1,a]{Sandeep Kaur}
\author[2,b]{Hitesh Sharma}
\author[3,c]{Vijay K. Jindal}
\author[4,5,d]{Vladimir Bubanja}
\author[6,e]{Isha Mudahar}
\affil[1]{Department of Physics, Punjabi University, Patiala, Punjab, India}
\affil[2]{Department of Applied Sciences, IKG Panjab Technical University, Kapurthala, Punjab, India}
\affil[3]{Department of Physics, Panjab University, Chandigarh, India}
\affil[4]{Measurement Standards Laboratory of New Zealand, Callaghan Innovation, PO Box 31310, Lower Hutt 5040, Wellington, New Zealand}
\affil[5]{The Dodd-Walls Centre for Photonic and Quantum Technologies, University of Otago, 730 Cumberland Street, Dunedin 9016, New Zealand}
\affil[6]{Department of Basic and Applied Sciences, Punjabi University, Patiala, Punjab, India}
\affil[a]{sandeep\_rs16@pbi.ac.in}
\affil[b]{hitesh@ptu.ac.in}
\affil[c]{vkjindal06@gmail.com}
\affil[d]{vladimir.bubanja@callaghaninnovation.govt.nz}
\affil[e]{dr.ishamudahar@gmail.com}
\date{}
\maketitle

\doublespace

\begin{abstract}
The geometric, electronic and magnetic properties of strained graphene nanoribbons were investigated using spin polarized calculations within the framework of density functional theory.  Cases of compressive stress along the longer axis of a nanoribbon and tensile stress at the midpoint and perpendicular to the plane of the nanoribbon were considered. Significant structural changes were observed including the formation of nanoripples. The calculated electronic and magnetic properties strongly depend on the size and shape of nanoribbons. The tunable magnetic properties of strained nanoribbons can be employed for designing magnetic nano-switches.
\end{abstract}

\section{Introduction}
\label{Introduction}
Graphene nanoribbons (GNRs) are quasi-one-dimensional nanostructures with unique electronic, optical and transport properties \cite{Chen1,Han2,Cresti3}. Therefore, they have been the subject of numerous studies considering their potential applications as field effect transistors, optoelectronic, spintronic and other devices \cite{Wang4,Jang,Soriano}. Of particular interest has been the possibility to tune the GNR properties by using the externally applied electric and magnetic fields, doping, defects, edge-modification and width-variation of nanoribbons \cite{HuangB,PanY,MajumdarK}. GNRs have also been incorporated in hybrid materials, such as vanadium dioxide-graphene architectures \cite{YangS}, exhibiting extraordinary electrochemical performance, and therefore showing promise for electrode material in high-power lithium batteries. After the original discovery of graphene \cite{Novoselov}, a plethora of other two-dimensional materials have been synthesised \cite{Xu} and have shown promise in a variety of fields, from photonics \cite{Xia} to nanomedicine \cite{Yang}. Calculations within the density functional theory framework have been carried out to investigate the electronic and magnetic properties of graphene-counterpart nanoribbons, such as those based on silicene \cite{Fang,Wang}, stenene \cite{Martin}, phosphorene \cite{Du}, and germanene \cite{Matthes,Monshi}, as well as to make first-principles predictions of new monolayer materials \cite{Chen}.

A range of top-down methods has been used to successfully fabricate GNRs. They include graphene etching \cite{Li}, chemical vapour deposition \cite{Pan}, scanning probe lithography \cite{Magda} and unzipping of single- \cite{Kosynkin} and multi-walled carbon nanotubes \cite{Jiao}. However, these techniques lack the synthetic reproducibility and atomically precise control required for predetermined electronic structure of GNRs. Instead, bottom-up approach can be used for this purpose.  A pioneering work \cite{Cai}, utilizing surface assisted coupling, has inspired a number of approaches based on appropriately designed precursor molecules to react in a selective way that ends up forming GNRs \cite{DiGiovannantonio,Sun}. This approach has been followed in solution \cite{Narita}, as well as supported on solid surfaces \cite{Talirz}.

Graphene has a remarkable stress-strain behaviour, including the highest stiffness and strength ever measured \cite{Lee,Bunch}. It can be easily bent to get complex folded structures \cite{KimK} and can withstand elastic deformations of up to 25 \% \cite{Lee}, that are much larger than in any other known material. Owing to these outstanding mechanical properties, graphene is an ideal candidate for nanomechanical systems \cite{Matthew} and flexible electronic devices \cite{Kim}. Measurements of the elastic response of graphene demonstrated that it is highly nonlinear for strains above 10 \% \cite{Lee}. Subsequently, these experiments were interpreted within a generalized nonlinear stress-strain relation that incorporates cubic terms in strain, and nonlinear elastic coefficients have been estimated from atomistic simulations \cite{Cadelano}.

When graphene is placed on a substrate, strains spontaneously arise due to either the lattice mismatch \cite{Ni}, or the surface corrugation of the substrate \cite{Teague}. Furthermore, intrinsic edge stress exists along the edges of graphene, causing warping instability \cite{Huang}.

Besides strain that is naturally generated, it can also be intentionally induced and controlled. Uniaxial strain can be induced by bending the flexible substrate \cite{Yu}. Biaxial strain can be introduced by utilizing the thermal mismatch between graphene and the substrate \cite{Ferralis}, using a bias voltage to shrink or elongate the piezoelectric substrate \cite{Ding}, or pushing the graphene clamped on top of a hole in the substrate with a tip of an atomic force microscope \cite{Lee}.

Numerous fascinating physical phenomena of graphene induced by strain have been reported, such as quantized pseudomagnetic field \cite{Levy}, zero-field quantum Hall effect \cite{Guinea}, enhanced electron-phonon coupling \cite{Si} and shifting of Dirac cones \cite{Pereira}.

While the band gap of graphene remains close to zero even for large strain, the band gap of GNRs is very sensitive to both uniaxial and shear strain. For an armchair GNR, uniaxial weak strain changes the band linearly, and for a large strain, it results in periodic oscillation of the band gap. On the other hand, shear strain always tend to reduce the band gap. For a zigzag GNR, the effect of strain is to change the spin polarization at the edges of ribbons modulating their band gap \cite{Lu}.

The planar structure of GNRs is deformed by formation of ripples due to an applied uniaxial strain \cite{Neek-Amal,WangZF,Guinea1}. Further, molecular dynamics studies have shown that the amplitude and orientation of ripples can be easily controlled by applied strain \cite{Baimova, Shenoy}. In addition, the spin polarized first principle calculations have shown that zigzag GNRs are magnetic in nature, whereas armchair GNRs show non-magnetic behaviour \cite{Owens}. When the strain is applied to the magnetic GNRs, magnetic moment is induced along zigzag edges with sinusoidal deformations \cite{Al-Aqtash}.

We have studied the effects of axial strain on structural, electronic and magnetic properties of armchair and zigzag graphene nanoribbons, as well as the effects of strain applied to the central atom of nanoribbons of square, triangular, rectangular and circular shapes.

\section{Computational Details}
All the calculations were performed using the Spanish Initiative for Electronic Simulation with Thousands of Atoms (SIESTA) computational code \cite{Soler}. The Perdew, Burkey and Ernzerhof (PBE) functional combined with double-zeta polarized basis set were used for the geometry optimizations \cite{Perdew}. The Kleinman-Bylander form of non-local norm conserving pseudo potentials were used to describe core electrons \cite{Kleinman}, while numerical pseudoatomic orbitals of the Sankey-Niklewski type \cite{Sankey} were used to represent the valence electrons. The size of pseudoatomic orbitals was defined by the energy shift parameter of 350 meV. The fineness of a grid was defined by the mesh cutoff of 250 Ry. The residual forces of the system were relaxed up to 0.03 eV/{\AA}. The ground state properties were found by minimizing the total energy of the system. The results of test calculations of the geometric and electronic properties of graphene and small fullerenes were found in agreement with the previous studies  \cite{Cooper,Kaur,Garg,SharmaH,SharmaA}.

\begin{figure}
\centering
\includegraphics[width=4in]{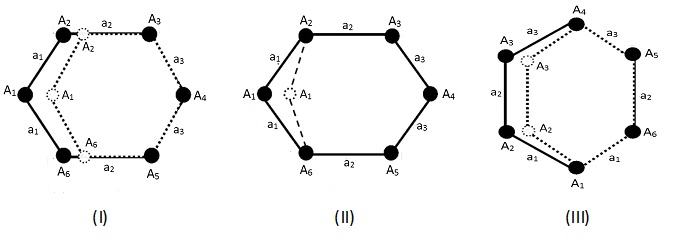}
\caption{Types of applied axial stress.}
\label{fig:strainimage}
\end{figure}

We have optimized ground state structures of GNRs for axial load applied at the two parallel edges of rectangular GNRs, as well as for a load applied at the center and perpendicular to a plane of square, triangular, rectangular and circular nanoribbons. We considered three types of boundary conditions in case of axial load, where either three, one or two atoms of the hexagons at the edges were shifted and remain fixed (as indicated in Figure \ref{fig:strainimage}). The atoms were moved inwards in steps of 0.2 {\AA} along the longer edge of 3 nm by 1 nm nanoribbon, which resulted in the compressive strain of nanoribbons in the range from 1.3\% to 6.7\%. In all three cases of axial loads, the free boundary conditions were imposed along the two unloaded edges. In case of the point load, we considered the effects of pulling the central C atom out of the nanoribbon plane by up to 5 {\AA}. For each setup considered, the ground state configurations and the binding energies per atom (B.E./atom) of GNRs were calculated. The spin polarized calculations were performed to obtain the density of states (DOS), the HOMO-LUMO gaps, the total magnetic moments (TMM) and the localized magnetic moments (LMM)of GNRs.

\section{Graphene nanoribbons under axial compression}
\begin{figure}
\centering
\includegraphics[width=5in]{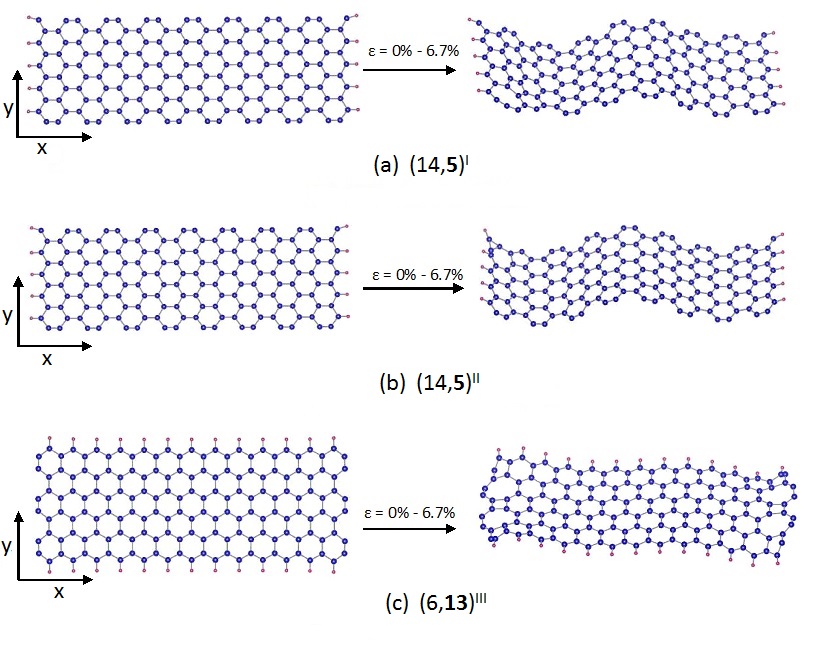}
\caption{Structure of nanoribbons; ;(a)$(14,\mathbf{5})^{I}$, (b) $(14, \mathbf{5})^{II}$, (c) $(6, \mathbf{13})^{III}$.}
\label{fig:nanoribbon1}
\end{figure}

\begin{figure}
\centering
\includegraphics[width=5in]{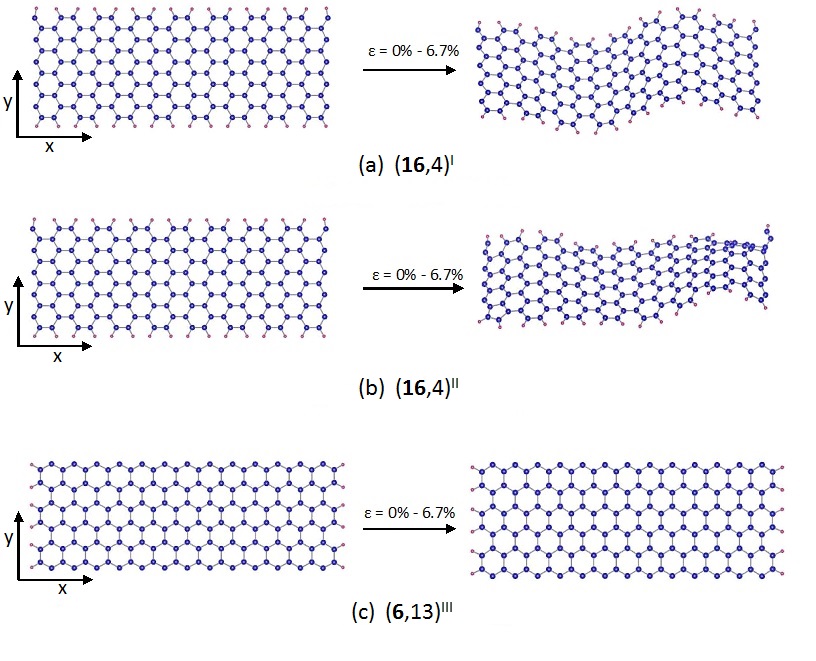}
\caption{Structure of nanoribbons; (a)$(\mathbf{16}, 4)^{I}$, (b) $(\mathbf{16}, 4)^{II}$, (c) $(\mathbf{6}, 13)^{III}$.}
\label{fig:nanoribbon2}
\end{figure}

\begin{figure}
\centering
\includegraphics[width=4in]{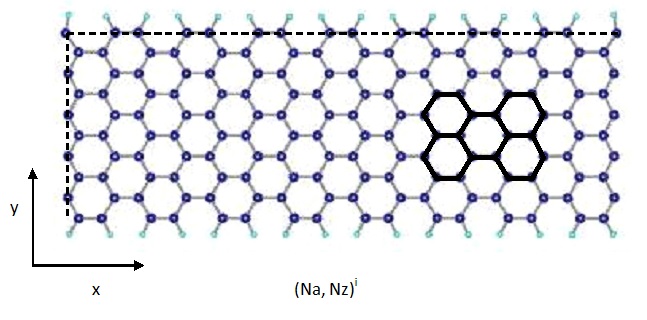}
\caption{Structure of GNR, where Na and Nz are counted along dashed lines and i is the type of applied stress. The highlighted area is shown in Figure 5.}
\label{fig:ribbon}
\end{figure}

In order to consider the edge effects on the properties of nanoribbons under axial compression, we considered the following six cases: $(14,\mathbf{5})^{(I)}$, $(14,\mathbf{5})^{(II)}$, $(6,\mathbf{13})^{(III)}$,  $(\mathbf{16},4)^{(I)}$, $(\mathbf{16},4)^{(II)}$ and $(\mathbf{6},13)^{(III)}$, where the notation $(N_a, N_z)^{(i)}$ denotes the nanoribbon with $N_a$ and $N_z$ carbon atoms along the armchair  and zigzag boundaries, respectively, under type-$i$ boundary conditions (see Figures \ref{fig:nanoribbon1} and \ref{fig:nanoribbon2}); the edges passivated by the hydrogen atoms are denoted in bold font (Figure \ref{fig:ribbon}).

\subsection{Structural properties}
\begin{figure}
\centering
\includegraphics[width=4in]{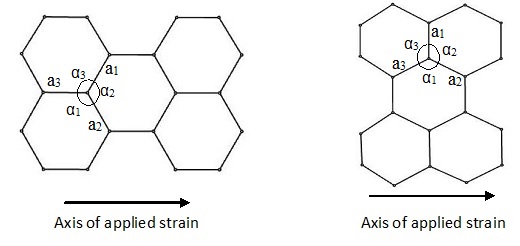}
\caption{Atomic structure of ribbons along armchair and zigzag edges, where $a_{1}$, $a_{2}$ and $a_{3}$ are bond lengths and $\alpha_{1}$, $\alpha_{2}$ and $\alpha_{3}$ are bond angles, respectively.}
\label{fig:typesofLA}
\end{figure}

\begin{figure}
\centering
\includegraphics[width=4in]{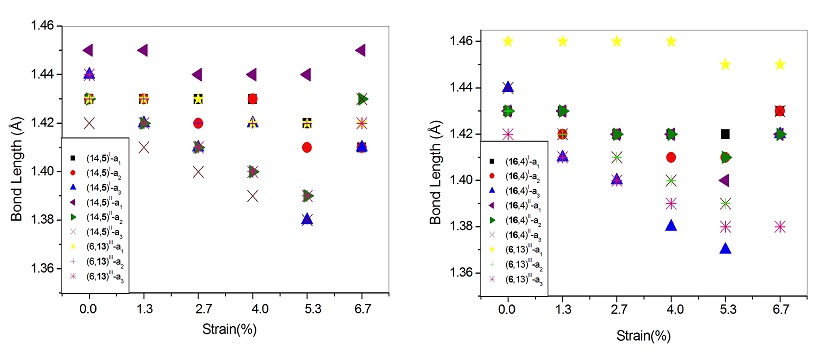}
\caption{Variation in bond lengths of nanoribbons with applied stress.}
\label{fig:bond}
\end{figure}

\begin{figure}
\centering
\includegraphics[width=4in]{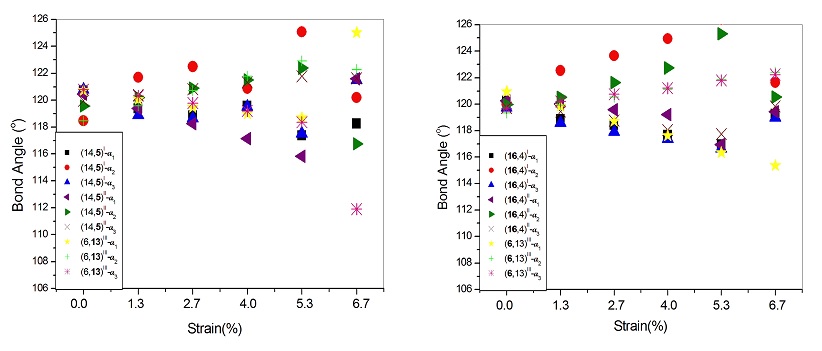}
\caption{Variation in bond angles of nanoribbons with applied stress.}
\label{fig:angle}
\end{figure}

The binding energies per atom of all considered GNRs are given in Table I, and show that configurations $(14,\mathbf{5})^{(I)}$, $(14,\mathbf{5})^{(II)}$, and $(\mathbf{6},13)^{(III)}$ are more stable than the other three. With the increase of strain the overall stability of GNRs is decreasing. To examine the geometry of the ground state structures we calculated the bond lengths and bond angles represented in Figure \ref{fig:typesofLA}. For a single vertex of a hexagon of a GNR, the variation in bond lengths and bond angles are given in Figures \ref{fig:bond} and \ref{fig:angle}, respectively. The bond lengths variations are of the order of 3 \%, whereas bond angles variations are of the order of 7 \%. For strain of up to 5.3\%, the bonds along the axis of the applied stress are shortened while those perpendicular to the applied stress are elongated. Above 5.3 \% strain, except for $(\mathbf{6}, 13)^{III}$ which remains planar, displacement of atoms along the vertical direction occurs, and the ripples are forming. GNRs start to follow generally sinusoidal shape, as expected from the continuum mechanics theory. The vertical displacement for the strain of 6.7 \% ranges from 1.3 {\AA} for $(6,\mathbf{13})^{(III)}$ to 2.4 {\AA} for $(\mathbf{16},4)^{(II)}$. Owing to the freedom of movement in the perpendicular direction to the GNR plane, the carbon atoms away from the boundaries tend to preserve the hexagonal shape of the lattice.  Near the boundaries some of the bonds are broken and octagonal shapes are formed.

\subsection{Electronic and magnetic properties}
\begin{figure}
\centering
\includegraphics[width=5in]{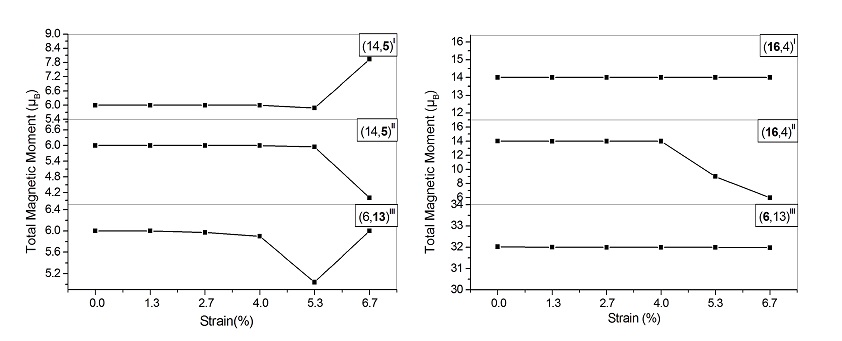}
\caption{Total Magnetic Moments for nanoribbons with applied axial stress.}
\label{fig:tmm}
\end{figure}

\begin{figure}
\centering
\includegraphics[width=5in]{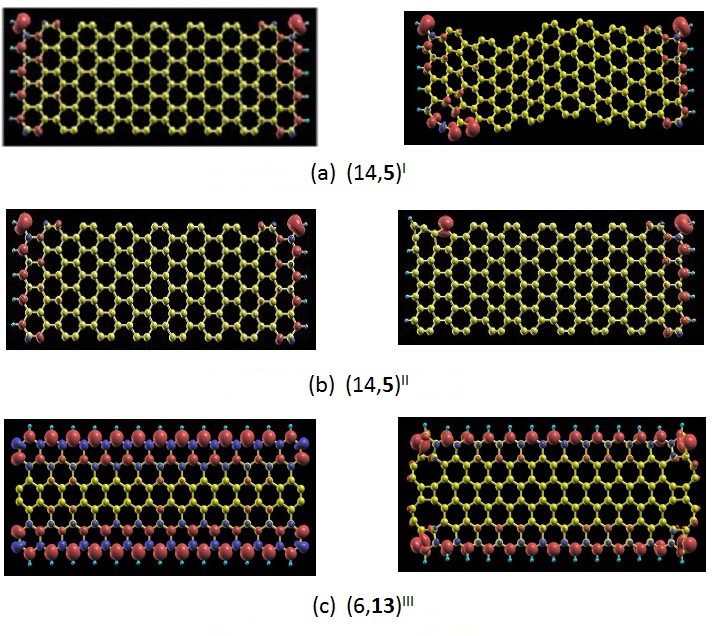}
\caption{Spin density maps for nanoribbons;(a)$(14,\mathbf{5})^{I}$, (b) $(14, \mathbf{5})^{II}$, (c) $(6, \mathbf{13})^{III}$.}
\label{fig:armchairlocal}
\end{figure}

\begin{figure}
\centering
\includegraphics[width=5in]{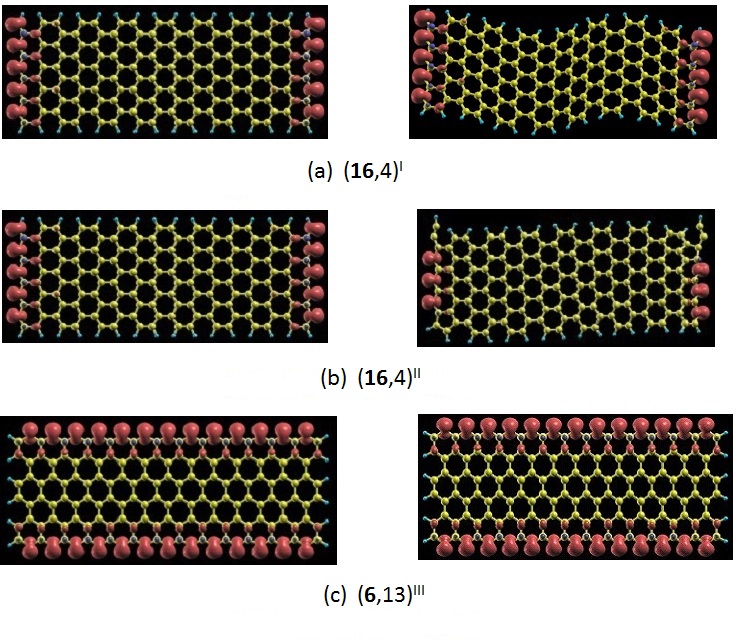}
\caption{Spin density maps for nanoribbons;(a)$(\mathbf{16}, 4)^{I}$, (b) $(\mathbf{16}, 4)^{II}$, (c) $(\mathbf{6}, 13)^{III}$.}
\label{fig:zigzaglocal}
\end{figure}

The total magnetic moments, as functions strain,  for all considered GNRs are shown in Figure \ref{fig:tmm}. It shows that TMMs change significantly from strain of 5.3 \%  for all except $(\mathbf{16},4)^{(I)}$ and $(\mathbf{6},13)^{(III)}$ . The magnetism in GNRs results from the states along the zigzag edges \cite{sahin}. When axial stress is applied, the structural distortions suppressing these states lead to a decrease of TMMs in $(\mathbf{14},5)^{(II)}$ and $(\mathbf{16},4)^{(II)}$, while in $(14, \mathbf{5})^{I}$ and $(6, \mathbf{13})^{III}$ the edge states arise, leading to an increase of TMMs. These observations can be clearly seen in Figures \ref{fig:armchairlocal} and \ref{fig:zigzaglocal}.

\begin{figure}
\centering
\includegraphics[width=5in]{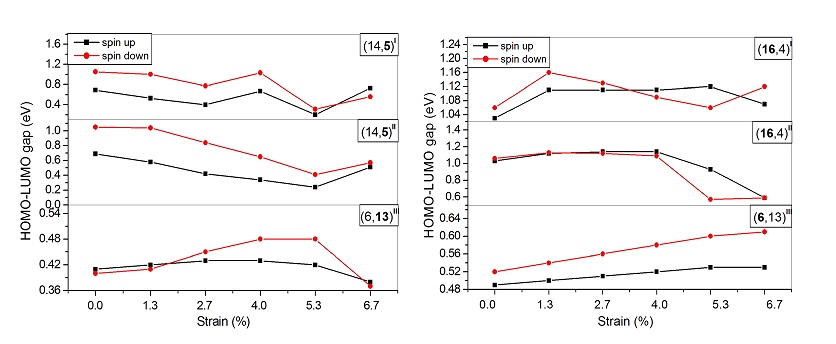}
\caption{HOMO-LUMO gaps in spin up and down states for nanoribbons with applied stress.}
\label{fig:gap}
\end{figure}

The HOMO-LUMO energy gaps for spin up as well as spin down electronic states are plotted in Figure \ref{fig:gap}. For $(\mathbf{6}, 13)^{III}$ GNR, the HOMO-LUMO gaps in both spin up and spin down states increase linearly with strain, while in other cases they vary nonuniformly. The finite energy difference between spin up and spin down HOMO-LUMO gaps confirm the magnetic behaviour of considered GNRs.

\begin{figure}
\centering
\includegraphics[width=6in]{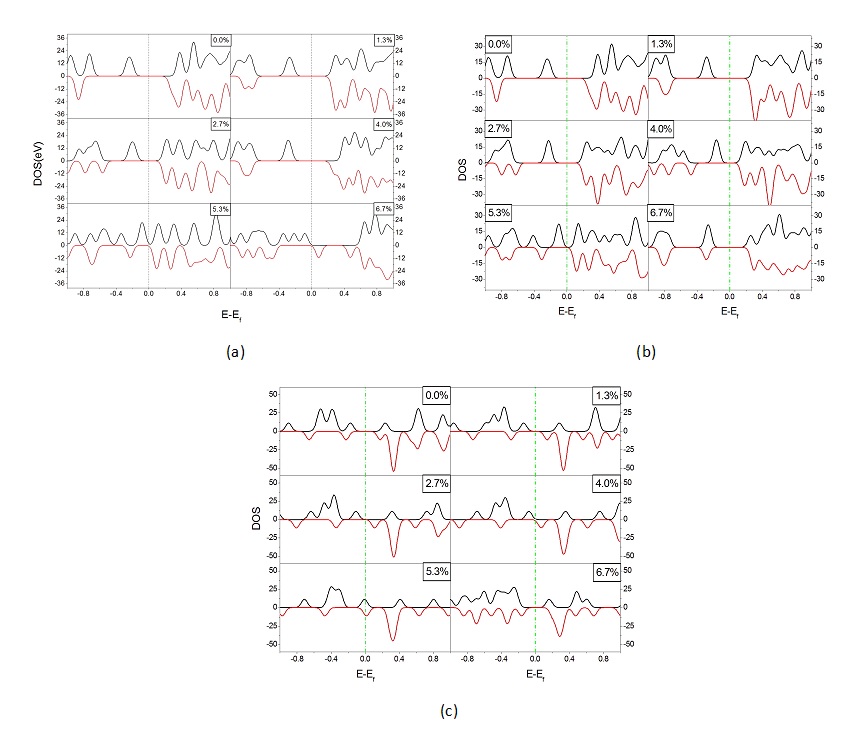}
\caption{Density of states for nanoribbons;(a)$(14,\mathbf{5})^{I}$, (b) $(14, \mathbf{5})^{II}$, (c) $(6, \mathbf{13})^{III}$. The black and red lines corresponds to DOS for spin up and spin down electronic states.}
\label{fig:dos}
\end{figure}

\begin{figure}
\centering
\includegraphics[width=6in]{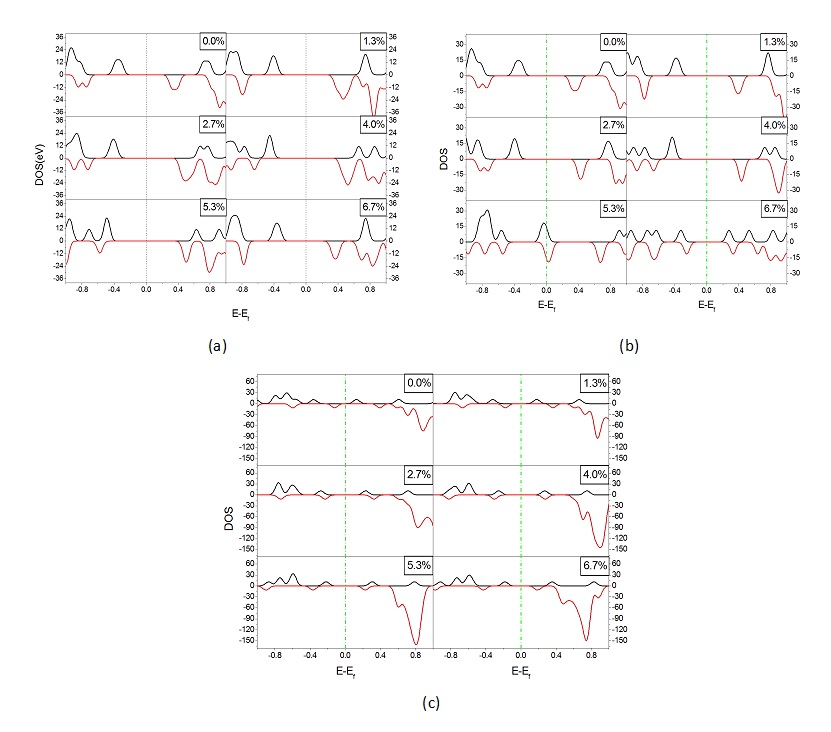}
\caption{Density of states for nanoribbons;(a)$(\mathbf{16}, 4)^{I}$, (b) $(\mathbf{16}, 4)^{II}$, (c) $(\mathbf{6}, 13)^{III}$. The black and red lines corresponds to DOS for spin up and spin down electronic states.}
\label{fig:dos1}
\end{figure}

The DOS plots (see Figures \ref{fig:dos} and \ref{fig:dos1}) show unequal distribution of spin up and spin down states, which leads to the magnetic behaviour of all GNRs. As the applied strain increases up to 5.3 \%, the energy states become available near the Fermi level, leading to an increase in conductivity. Above 5.3\% strain, GNRs conductivity decreases.

\section{Graphene nanoribbons under concentrated load}
\subsection{Structural properties}
\begin{figure}
\centering
\includegraphics[width=4in]{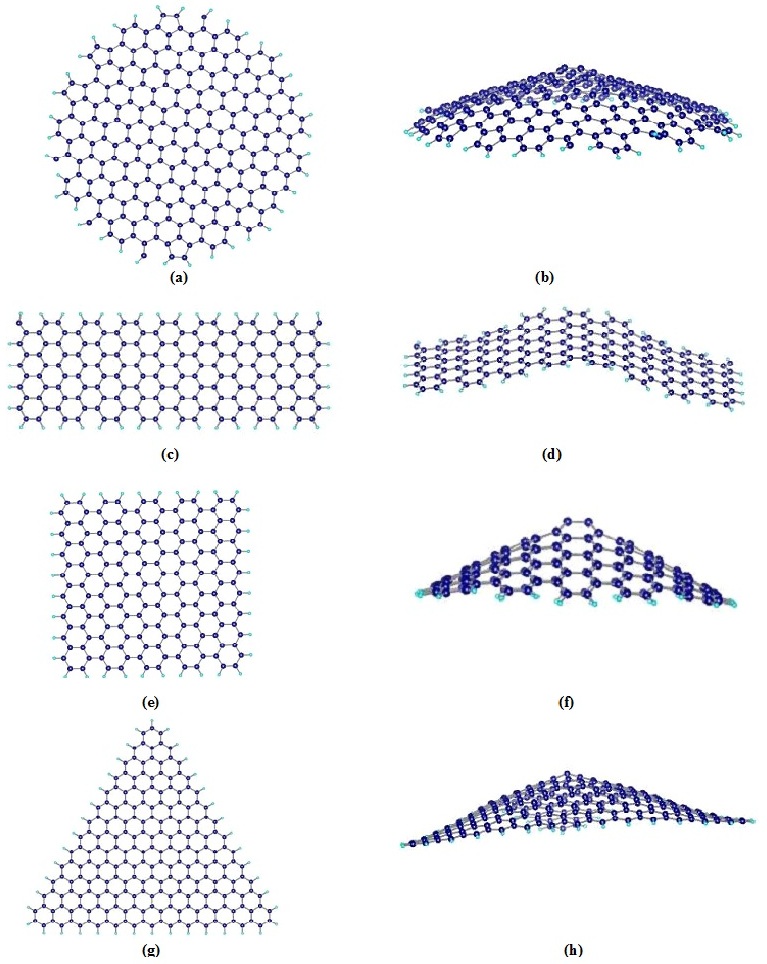}
\caption{Graphene naoribbons before [(a),(c),(e)] and after [(b),(d),(f)] applied point stress.}
\label{fig:different}
\end{figure}

We considered GNRs of square, triangular, rectangular and circular shapes under point stress applied by pulling the middle carbon atom out of the nanoribbon plane by 5 {\AA}. The pristine and relaxed structures after applying the stress are shown in Figure \ref{fig:different}. The corresponding structural parameters are shown in Table \ref{different}. The calculated bonding energies per atom decrease as the stress is applied for all shapes except square, in which case the binding energy remained the same. The C-C bond lengths of rectangular ribbon lie in the range 1.37-1.46 {\AA} for pristine as well as strained GNR. However C-C bond lengths increase significantly as functions of the applied strain for the other three GNRs (Table \ref{different}). The bond angles for rectangular ribbon experience slight variations with increase in strain, whereas for the other three GNRs, the bond angles significantly vary.

\subsection{Electronic and magnetic properties}
\begin{figure}
\centering
\includegraphics[width=5in]{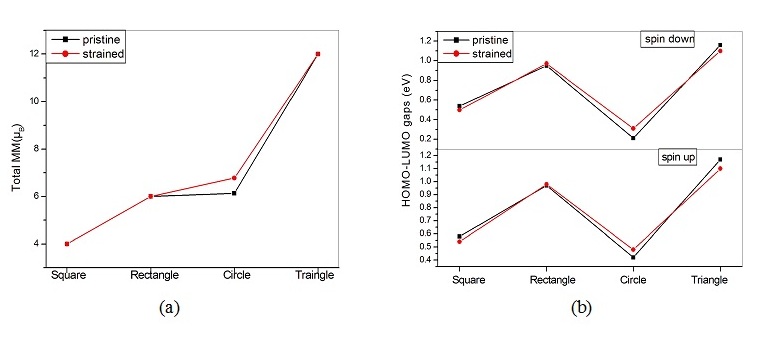}
\caption{Total Magnetic Moments and HOMO-LUMO gaps for nanoribbons with applied point stress.}
\label{fig:differentgm}
\end{figure}

The spin polarized calculations are performed for all four differently shaped GNRs. The TMMs and HOMO-LUMO gaps for spin up and spin down states are shown in Figure \ref{fig:differentgm}. The TMM increases with the application of strain for the circular shaped GNR from 6.13 $\mu_B$ to 6.78 $\mu_B$, while in the other three cases it remains the same.

\begin{figure}
\centering
\includegraphics[width=5in]{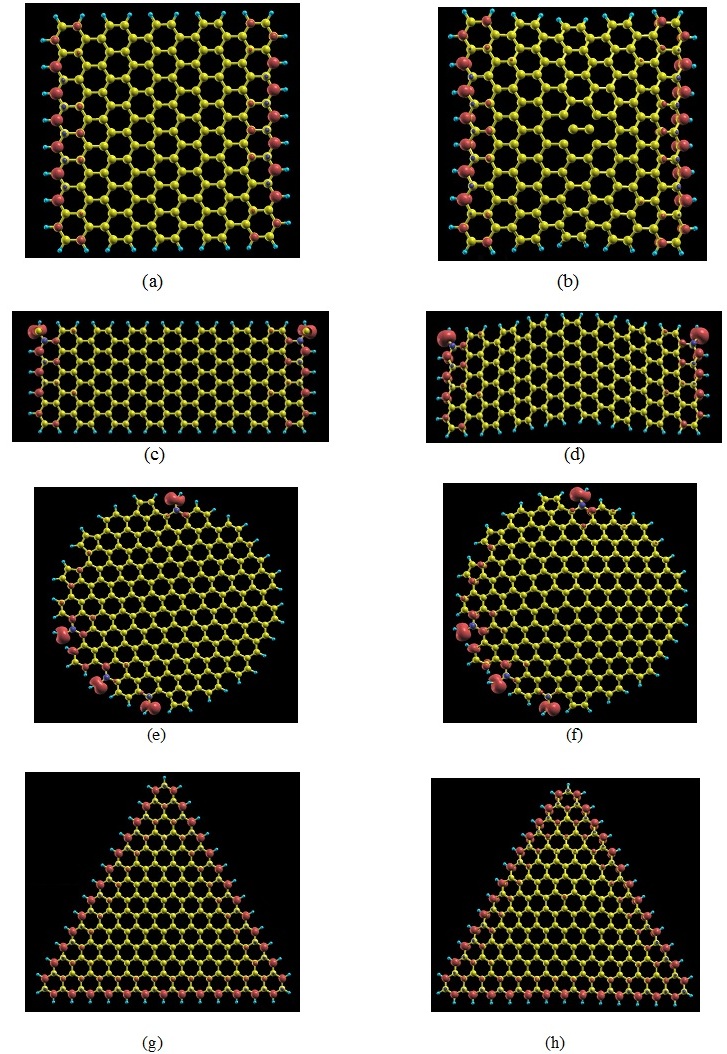}
\caption{Spin density maps for nanoribbons before [(a),(c),(e),(g)] and after [(b),(d),(f),(h) applied point axial stress.}
\label{fig:differentlocal}
\end{figure}

The localized MMs are presented in Figure \ref{fig:differentlocal}, showing that there is no local MM at the central C site under strain for the square, rectangular and circular ribbons, while in case of the triangular shaped ribbon, localized MM is induced at the central atom of the ribbon. The major contribution towards TMM comes from the edge atoms. The HOMO-LUMO gaps for both pristine and strained nanoribbons show similar pattern in both spin up and spin down electronic states. There is an increase in HOMO-LUMO gaps with applied strain for circular and rectangular nanoribbons, whereas for square and triangular ribbons, the HOMO-LUMO gaps decrease after introduction of strain.

\begin{figure}
\centering
\includegraphics[width=5in]{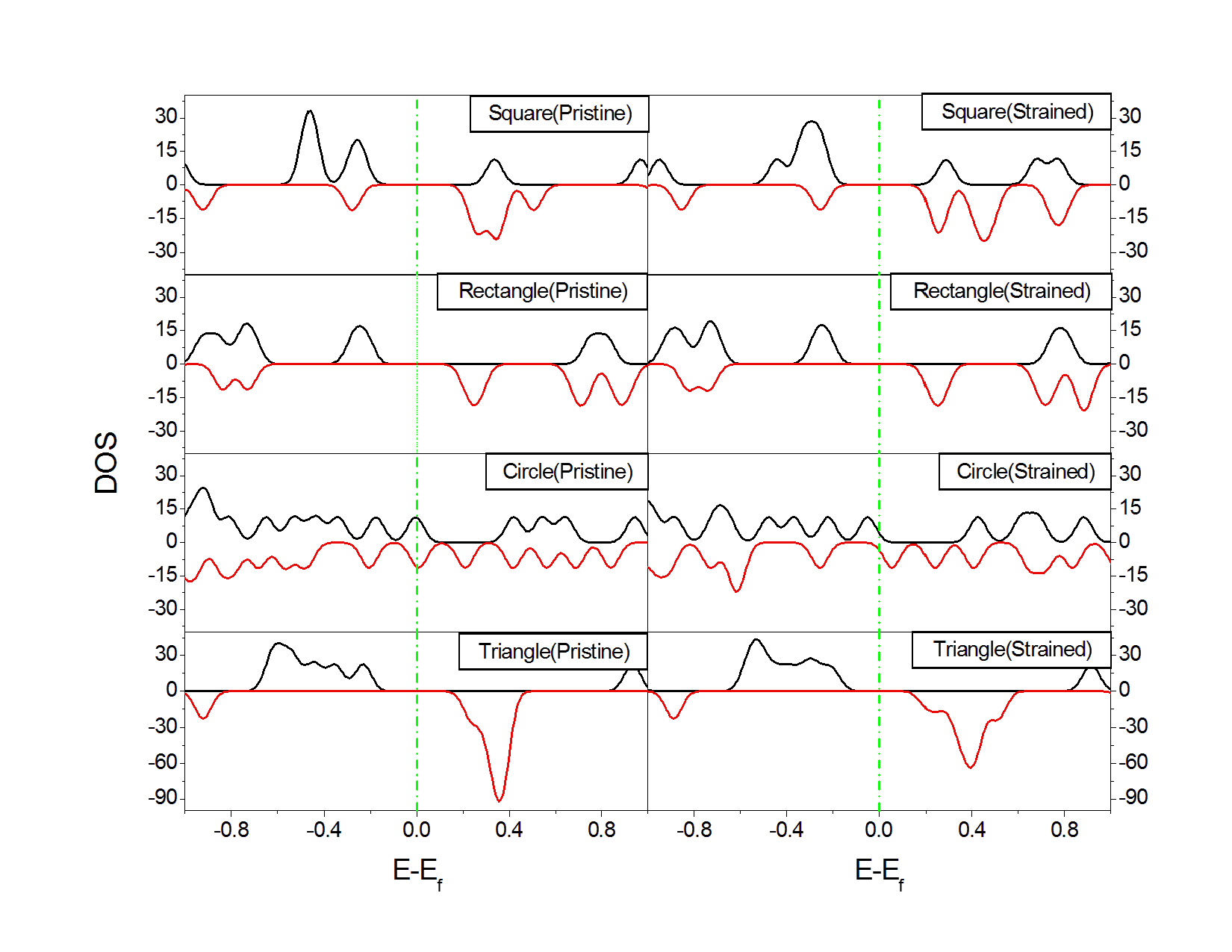}
\caption{Density of states for nanoribbons with applied point stress. The black and red lines corresponds to DOS for spin up and spin down electronic states.}
\label{fig:differentdos}
\end{figure}

Density of states for the considered four shapes of graphene nanoribbons are shown in Figure \ref{fig:differentdos}. Plots confirm the magnetic behaviour of all the four shapes of GNRs as a result of unequal spin density states. DOS plots also show that there is a redistribution and spin polarization of electrons in spin up and down states as the point stress is applied.

The above considerations show that the structural, magnetic and electronic properties of graphene nanoribbons are shape dependent and can be tuned by the  applied point strain, which makes them potentially useful in applications involving spintronic devices and stress sensors.

\section{Conclusions}
The spin polarized density functional theory has been employed to study the structural, electronic and magnetic properties of strained graphene nanoribbons. We considered the cases of axial compressive stress applied at two parallel edges of rectangular nanoribbons and point stress applied at the middle of nanoribbons of square, rectangular, circular and triangular shapes. With the application of axial strain, the C-C bond angles start changing, and significant rippling is observed at 6.7 \%. Near the edges where the stress is applied, octagonal shapes of carbon atoms were formed replacing the original hexagons.

The electronic and magnetic properties of graphene nanoribbons also change with the introduction of axial stress. The total magnetic moment shows a significant variation with the application of stress. The variation in TMM is directly proportional to distortion in hexagonal lattice and the number of zigzag edge states. The spin density maps show the imbalance of spin up and spin down states which gives rise to the magnetic behaviour of nanoribbons. The finite energy difference between HOMO-LUMO gaps for spin up and spin down states also confirms the magnetic behaviour of GNRs. The change in DOS near Fermi level, creating unoccupied states, takes place with increase in strain.

The application of a point stress at the centre of ribbons induces magnetism at the centre of triangular shaped GNRs, whereas for the other ribbons localized moments only lie at the edge atoms. The calculated magnetic properties show that stress induces significant changes which could be employed in spintronic applications. Rippling effects in GNRs may find applications in mechanical sensors.

\section*{Acknowledgements}
The authors are grateful to the New Zealand eScience Infrastructure (NeSI) for providing high performance computing facilities. The authors are also thankful to the UGC (University Grant Commission)-New Delhi, India for their financial support.

\newpage
\bibliographystyle{plain}

\newpage
\begin{table}[]
\caption{Binding energies per atom in eV  for graphene nanoribbons under axial compression}
\label{label}
\begin{tabular}{|l|l|l|l|l|l|l|}
\hline
 Strain (\%)& $(14,\mathbf{5})^{(I)}$ & $(14,\mathbf{5})^{(II)}$ & $(6,\mathbf{13})^{(III)}$ & $(\mathbf{16},4)^{(I)}$ & $(\mathbf{16},4)^{(II)}$  & $(\mathbf{6},13)^{(III)}$  \\ \hline
0.0 & 8.30 & 8.30 & 7.87 & 7.63 & 7.63 & 8.23 \\ \hline
1.3 & 8.29 & 8.30 & 7.87 & 7.61 & 7.62 & 8.22 \\ \hline
2.7 & 8.27 & 8.28 & 7.86 & 7.60 & 7.61 & 8.21 \\ \hline
4.0 & 8.28 & 8.27 & 7.84 & 7.58 & 7.60 & 8.19 \\ \hline
5.3 & 8.22 & 8.24 & 7.81 & 7.56 & 7.57 & 8.16 \\ \hline
6.7 & 8.27 & 8.28 & 7.76 & 7.61 & 7.62 & 8.11 \\ \hline
\end{tabular}
\end{table}

\begin{table}
 \caption{Structural parameters for graphene nanoribbons under point stress}
 \label{different}
 \begin{center}
  \begin{tabular}{|c|p{1.5cm}p{2cm}p{2cm}|p{1.5cm}p{2cm}p{2cm}|}
\hline
    Ribbon & & Pristine & & & Strained & \\
\hline
    Shape & B.E./atom (eV) & Bond lengths(\AA) & Bond angles($^{\circ}$) & B.E./atom (eV) & Bond lengths(\AA) & Bond angles($^{\circ}$) \\
\hline
    Rectangle  & $7.42$ & $1.37-1.46$ & $117.89-123.12$ & $7.42$ & $1.38-1.46$ & $117.83-123.02$ \\
    Square & $7.68$ & $1.38-1.45$ & $118.08-121.83$ & $7.61$ & $1.35-1.65$ & $102.67-124.36$ \\
    Triangle & $7.81$ & $1.40-1.44$ & $118.73-122.38$ & $7.79$ & $1.39-1.54$ & $110.97-122.92$ \\
    Circle & $7.92$ & $1.35-1.48$ & $111.73-125.35$ & $7.88$ & $1.34-1.56$ & $108.63-126.28$ \\
\hline
	\end{tabular}
 \end{center}
\end{table}

\end{document}